# Exceptionally accurate large graphene quantum Hall arrays for the new SI

Hans He[1], Karin Cedergren[1], Naveen Shetty[2], Samuel Lara-Avila[2,3], Sergey Kubatkin[2], Tobias Bergsten[1] and Gunnar Eklund[1].

1: RISE Research Institutes of Sweden, Box 857, S-50115 Borås, Sweden

2: Department of Microtechnology and Nanoscience, Chalmers University of Technology, 412 96 Gothenburg, Sweden

3: National Physical Laboratory, Hampton Road, Teddington TW11 0LW, United Kingdom

**The quantum Hall effect (QHE) is a cornerstone in the new International System of Units (SI), wherein the base units are derived from seven fundamental constants such as Planck's constant $h$ and elementary charge $e$[1]. Graphene has revolutionized practical resistance metrology by enabling the realization of quantized resistance $h/2e^2 \approx 12.9$ k$\Omega$ under relaxed experimental conditions[2–4]. Looking ahead, graphene also has the potential to improve realizations of the electronic kilogram using the Kibble balance[5], and the quantum Ampere in wide current ranges[6,7]. However, these prospects require different resistance values than practically achievable in single QHE devices, while also imposing stringent demands on energy dissipation in single QHE devices, ultimately requiring currents almost two orders of magnitude higher than the typical QHE breakdown currents $I_c \sim 100$ μA achievable in graphene[3,4,8]. Here we present unprecedented accuracy in the quantization of a record sized quantum Hall array (QHA), demonstrating $R_K/236 \approx 109$ $\Omega$ with 0.2 part-per-billion (n$\Omega/\Omega$) accuracy with $I_c \geq 5$ mA (~ 1 n$\Omega/\Omega$ accuracy for $I_c = 8.5$ mA), using epitaxial graphene on silicon carbide (epigraphene). The array quantization accuracy, comparable to the most precise universality tests of QHE in single Hall bar devices[9,10], together with the scalability and reliability of this approach pave the road for superior realisations of three key units in the modern SI: the ohm, the ampere, and the kilogram.**

Epigraphene QHE devices are the preferred embodiment of the quantum Hall primary electrical resistance standard, providing an exact relationship between resistance and fundamental constants $R = R_K/4(N+1/2)$, expressed using the Von Klitzing constant $R_K = h/e^2$ and an integer $N \geq 0$. Epigraphene combines a large Landau level energy gap, typical of Dirac fermions in graphene[11], with high energy loss rates (i.e. high electron phonon coupling) that result in larger $I_c$ compared to conventional semiconductors[12,13]. Moreover, the large quantum capacitance of the epigraphene system leads to a magnetic field dependent charge transfer from the SiC substrate, which results in the widest quantum N = 0 resistance plateau observed to date, extending over 50 T[14,15]. The N = 0 plateau is not only the most robust, but also the most well-quantized and is therefore preferred for precision metrology[2–4]. In practice, all of these material-specific virtues translate into highly robust quantization over a wide parameter space[3] and greatly facilitates practical quantum resistance metrology.

In the 2019 redefinition of the SI, the QHE is gaining more prominence due to its elevation from practical to true realization of resistance, and it will serve other roles beyond resistance calibration. One such exciting application is the realisation of the electronic kilogram via the Kibble balance, which in a nutshell determines the weight of the object in terms of a measured current $I$ and voltage $V$. While $V$ can be measured by comparing it to a primary Josephson voltage standard[16], the current measurement still relies on a secondary artefact resistor which has to be calibrated against the QHE in a separate step. The direct integration of a QHE primary resistor in the Kibble balance could increase its performance, while also decreasing the complexity of the measurements. Such a feat would require a devices with resistance and $I_C$ on the order of 100 Ω and 10 mA respectively[17]. Furthermore, if QHE devices with arbitrary resistance and high $I_C$ could be implemented, they could using Ohm's law be combined with existing programmable Josephson array voltage standards to realize the quantum ampere over ranges far beyond current pumps[18], and without high external amplification[6,7]. Moreover, QHE devices with different resistances are also immensely useful for practical resistance metrology and will reduce uncertainties in calibration of a wide range of resistance values. However, a technological breakthrough is needed to enable the aforementioned applications, since a single graphene Hall bar can in practice only achieve $R = R_K/2$ and $I_C \sim 100$ μA at typical operating conditions[2,3,8].

The use of arrays of quantum Hall bars is an elegant way to provide in principle quantized resistance at arbitrary levels via series and parallel connections of individual Hall devices[19–23], while effectively increasing $I_C$ via parallel resistances. However, QHAs have not until now truly met the stringent criteria required for a metrological standard in terms of precision and reliability. A great challenge associated with the QHA endeavour is to achieve a 100% device yield. Any minor imperfection in any one individual Hall bar, be it improper quantization or poor contact resistance, will be detrimental to quantization accuracy. In practice, this implies that achieving sub part-per-billion accuracy requires that the combined effects of contacts, wiring and residual longitudinal resistance $R_{xx}$ should be less than 100 nΩ for a QHA with 100 Ω resistance. Another unresolved issue is associated with the measurements of vanishing $R_{xx} \sim 0$, which is an established test of resistance quantization[24]. While $R_{xx}$ can be assessed in individual Hall bars one at a time, this approach is not feasible for large-scale arrays.

Here we present quantum Hall measurements performed on a record-size QHA device, with 236 individual epigraphene Hall bars. We propose that a direct comparison between two epigraphene QHAs, analogous to QHE universality tests between GaAs and graphene[10], is the best method to verify the accuracy of quantization, circumventing the need to measure $R_{xx}$ in each array element. We demonstrate that a direct comparison between two large epigraphene arrays using high-precision measurements show no significant deviation of their resistance within 0.2 nΩ/Ω, with mutual agreement comparable to the best universality tests of QHE to date[9]. Our measurements are further validated through additional comparisons between the array, a single epigraphene quantum Hall bar, and a secondary 100 Ω resistance standard.

The array contains 236 individual Hall bars (Fig. 1a), divided between two subarrays (Array1 and Array2) connected in series, each with 118 Hall bars in parallel and a nominal resistance of $h/236e^2$ = 109.376302794 Ω (whole array R = $h/118e^2$) at the N = 0 plateau. The Hall bars are circular in order to achieve symmetrical design with high packing density. To maximize $I_C$, the diameter was chosen to be 150 μm so that the distance between contacts exceeds the equilibration length of the QHE edge state which is on the order of 100 μm at 2K and 5 T[25]. The contacts and interconnects were made from superconducting niobium nitride (NbN) (Supplementary 1), and were dimensioned to be at least 120 nm thick and 50 μm wide to support currents on the order of 10 mA at 2 K and 5 T[26]. The NbN is in direct contact with epigraphene, with a split contact design using six connections to minimize the contact resistance[27]. The carrier density was tuned using molecular doping[28], which reliably yields low charge disorder and proper quantization, and stability over years[8]. The array exists on the same chip together with individual Hall bars, and all measurements were performed in the same cryostat and using the same setup. The proximity of the devices minimizes external influences due to excess wiring, and the direct one-to-one ratio comparison between the subarrays further reduces many uncertainty contributions and errors in the precision measurements. Devices were tested simultaneously by performing a direct comparison of their quantized resistance values via a cryogenic current comparator (CCC) system, which is a well-established method to measure resistance ratios with the highest precision[2,3,8]. It can detect minute deviations $\Delta$ from 100 $\Omega$ on the order of 10 n$\Omega$ (0.1 n$\Omega/\Omega$)[29] and makes for the ultimate test of resistance quantization.

Figure 1 contains the main results of this work: the mean relative deviation of the direct subarray comparison demonstrating that the resistance of each subarray is the same within 0.2 n$\Omega/\Omega$. Each data point in Fig. 1b is the weighted mean of multiple CCC-readings (≥ 45 readings), each taking 20 min and consisting of multiple current polarity shifts to compensate thermal voltages and short-term drift. The standard deviation of each reading is used as a weight in the calculation of the final mean (see Methods). Allan deviation analysis is used to characterize the type of noise present in the measurement[30]. It decreases with elapsed measurement time $\tau$ as ~ $1/\tau^{1/2}$ (Fig. 1c) indicating that white noise is the dominating type. It also shows that the minimum measured uncertainty for the standard error in our experiments is in practice 0.2 n$\Omega/\Omega$. A histogram (Fig. 1d) shows that the data used in the above analysis are normally distributed and further supports the notion that white noise dominates.

The weighted mean of the mean relative deviations $\Delta_{Array1-Array2}$ at different fields in Fig. 1b reveals the level of quantization[3,9,21]. Using the standard error as the weight (see Methods), the resulting weighted mean relative deviation and standard error of the weighted mean is $\Delta_{Array1-Array2}$ = (0.033 ± 0.082) n$\Omega/\Omega$. This degree of accuracy in the quantization of such a large QHA is unprecedented for both GaAs[23] and graphene[20,22], and it is well below 1 n$\Omega/\Omega$ which is the requirement for precision metrology[24]. Moreover, this result is comparable to the most accurate comparisons of single graphene Hall bars versus GaAs in universality tests of QHE, which used the same analysis and reported a deviation of $\Delta_{GaAs-Graphene}$ = (-0.047 ± 0.086) n$\Omega/\Omega$[9]. Note however that due to the higher measured Allan deviation, the metrological confidence is valid for uncertainties down to 0.2 n$\Omega/\Omega$, limited by our setup.

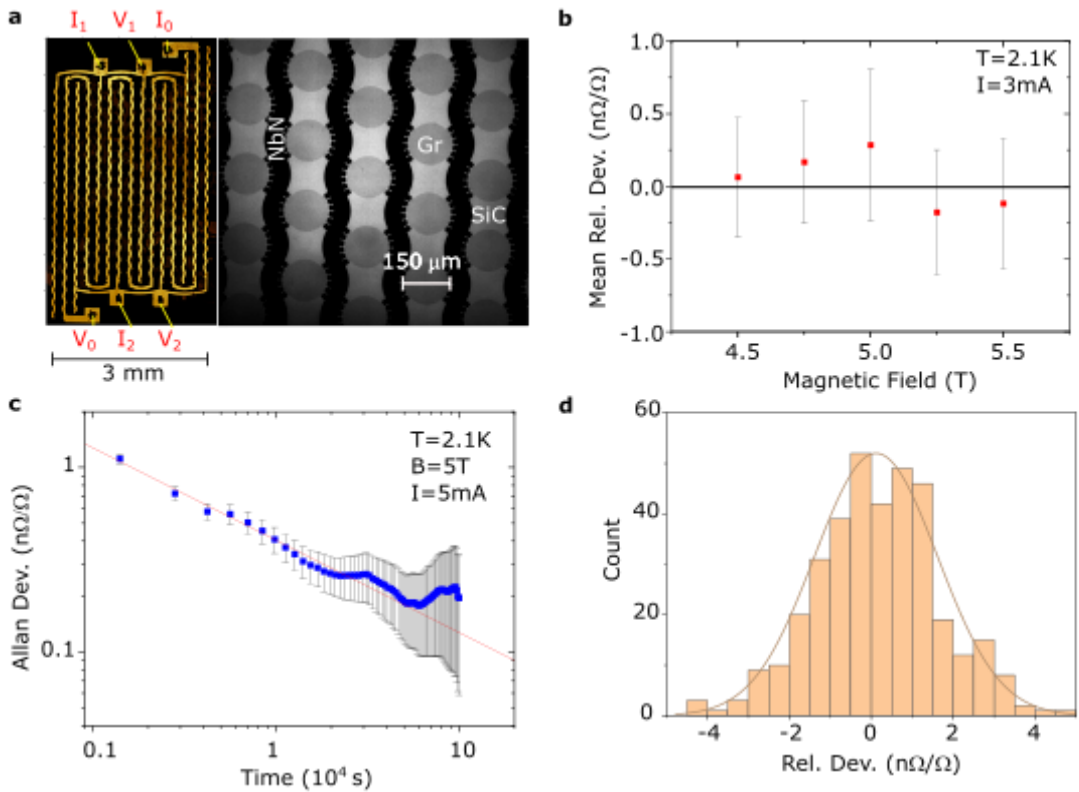

*Figure 1. **Direct comparison measurements of epigraphene arrays**. **a,** Left image is a false color composite micrograph of the whole array. It consists of two subarrays connected in series, each with 118 Hall bars in parallel for a total of 236 Hall bars. Subarray 1 and Subarray 2 are biased and measured using superconducting NbN leads connected to $I_1$-$I_0$ and $V_1$-$V_0$, and $I_2$-$I_0$ and $V_2$-$V_0$ respectively. The right image shows a zoomed-in transmission mode micrograph of the individual circular graphene Hall bars, which are connected in a simple two-probe configuration using split contacts. **b,** Precision CCC-measurements taken at different magnetic fields show the mean relative deviation between the two subarrays. Each point is the weighted mean of ≥ 45 CCC-readings, each around 20 minutes long, and the error bars represent one standard error derived from Allan deviation at $10^4$ s. These measurements reveal that there is no significant deviation over the measured field range **c,** Allan deviation follows $1/\tau^{1/2}$ (red line) which indicates that white noise dominates and limits the measurement uncertainty to 0.2 n$\Omega/\Omega$. At longer averaging times the Allan deviation no longer decreases due to 1/f noise and drift dominating. The error bars are estimated relative errors. **d,** Histogram of the data which produced the means shown in **b**. Each count represents one 20 minutes long CCC-series. The distribution is normal, and the solid line is a Gaussian fit which shows that the unweighted mean lies around 0.1 n$\Omega/\Omega$.*

This level of agreement between the resistance of the subarrays can only be attributed to exact quantization. Especially because the subarrays, though nominally identical, are expected to have slightly different non-quantized resistance due to finite doping difference (Supplementary 2). We have also compared the subarrays to an on-chip single Hall bar, in order to further verify the quantization and to form a link between our measurements and traditional quantum Hall experiments[24]. The Hall bar was dimensioned to be 200 μm wide, comparable to an individual array element, so that their $I_C$ are similar. The Hall bar characterization (Fig. 2a) shows that its longitudinal resistance $R_{XX}$ = $V_X$/I vanishes into the noise level of ~ 100 nV (limited by setup) above the quantizing field $B$ = 3 T, same as for the array (Supplementary 1). Fig. 2b shows the bias current dependence of $R_{XX}$ of the Hall bar has no significant change up to 100 μA, and the $I_C$ for the Hall bar is therefore around 100 μA. This also suggests that the $I_C$ of an individual array Hall element should be on a similar level. The mean residual $R_{XX}$ for bias currents 5-100 μA is $R_{XX}$ = (0.2 ± 0.1) m$\Omega$, which approaches zero within the noise for two standard deviations. A residual resistance of 0.1 m$\Omega$ could lead to a deviation of the quantized resistance $h/2e^2$ on the order of 3 n$\Omega/\Omega$[3], and would be easily identified in CCC-measurements. The contact resistances (same NbN split contacts as array) were measured under quantizing conditions using a standard 3-probe configuration[24] and were all < 2 $\Omega$, including ~1.5 $\Omega$ wire resistance, well-below recommended levels[24]. In summary, the Hall bar passed all established tests for initial characterization of a single Hall bar resistance standard.

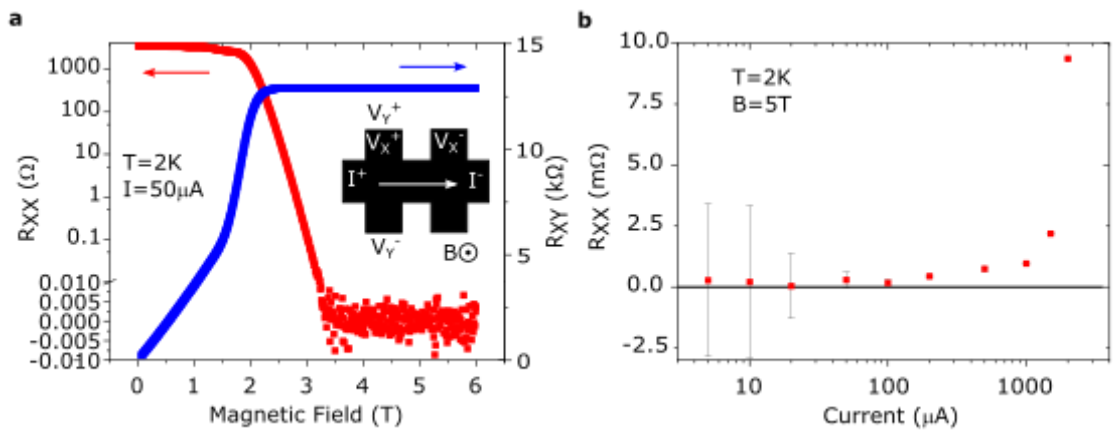

*Figure 2. **Characterization of single Hall bar**. **a,** A separate Hall bar with a normal rectangular geometry is used to measure $R_{XX}$ (red) and $R_{XY}$ (blue). The device is fully quantized for B > 3 T, and the longitudinal resistance vanishes below the noise level of ~ 100 nV. From the low-field measurement of the transverse resistance $R_{XY}$ = $V_Y$/I, the carrier density is n = $1.7 \times 10^{11}$ cm$^{-2}$ and mobility is μ = 19,600 cm$^2$/Vs. Since all quantum Hall devices are*

*located on the same chip, this also provides an indirect measurement of the array carrier density and mobility.* ***b****, Critical current measured on the Hall bar reveals no significant increase in $R_{xx}$ up to 100 μA bias. Error bars represent one standard deviation.*

Fig. 3a shows the comparison between the Hall bar and a 100 Ω standard resistor, and each subarray versus the same 100 Ω standard. The 100 Ω standard is kept immersed in a temperature-controlled oil bath, with a well-recorded history and long-term stability. An indirect comparison between the Hall bar and subarrays using these data results in the deviation and combined uncertainty of $\Delta_{HB-100}$ - $\Delta_{Array1-100}$ = $\Delta_{HB-Array1}$ = (-0.2 ± 1.9) nΩ/Ω and $\Delta_{HB-Array2}$ = (0.1 ± 2.0) nΩ/Ω. The combined uncertainty is dominated entirely by the Hall bar measurement, which is noisier because the ratio of $h/2e^2$ over 100 Ω is far from unity, and therefore more sensitive to noise in the CCC-balance. The indirect comparison between subarrays is $\Delta_{Array1-100}$ - $\Delta_{Array2-100}$ = $\Delta_{Array1-Array2}$ = (0.3 ± 0.6 nΩ/Ω), in good agreement with the direct array comparison.

To add further confidence to our measurements, we also performed a direct comparison of the Hall bar and one subarray. Fig. 3b shows one long CCC measurement, including Allan deviation. Fig. 3c shows similar measurements taken at different fields and Fig. 3d shows that the data across all fields is dominated by white noise. Taking the weighted mean of all points (same as for data in Fig. 1b), the calculated mean deviation for the direct comparison is $\Delta_{HB-Array1}$ = (-0.04 ± 0.2) nΩ/Ω, in good agreement with the direct subarray comparison. We have now demonstrated agreement between different combinations of direct and indirect comparisons between a quantized standard Hall bar, a 100 Ω standard, and the subarrays, and the measured deviations are all consistent with each other (Supplementary 3).

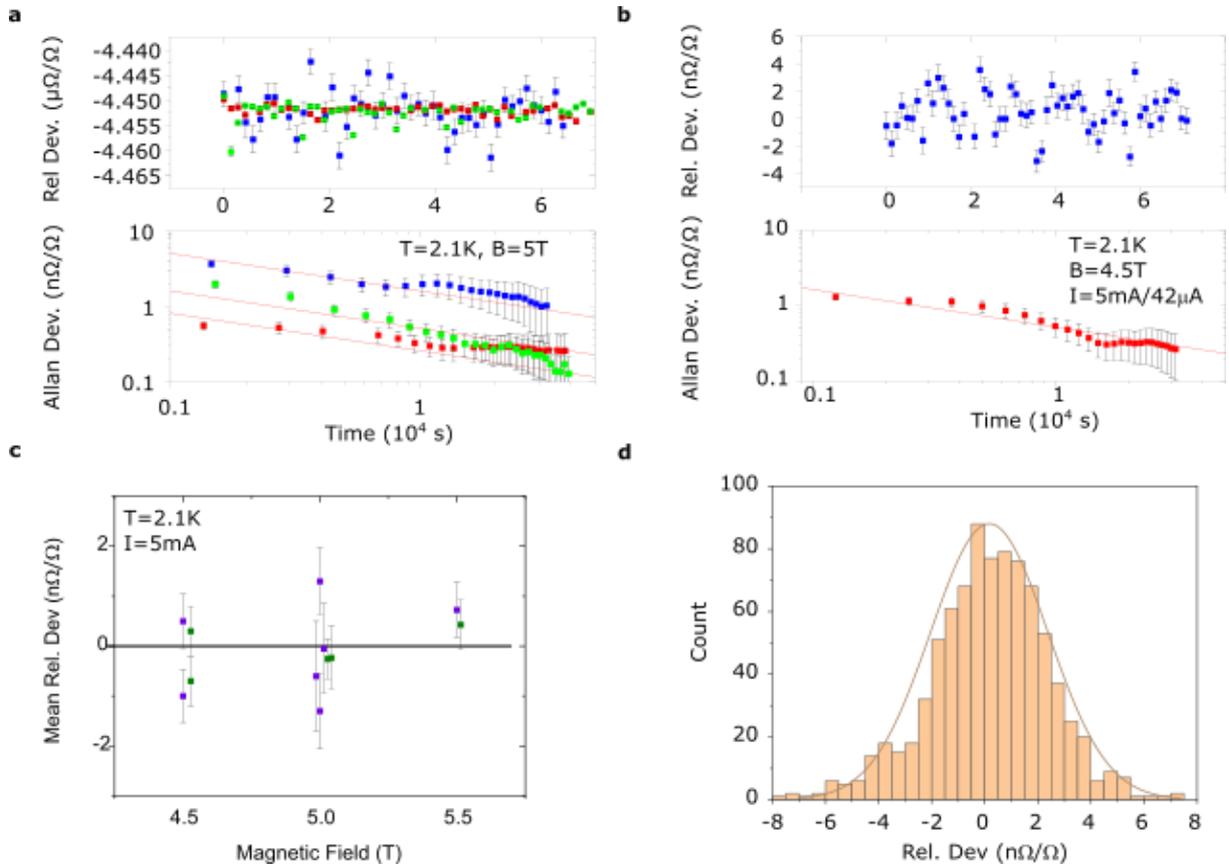

*Figure 3.* **Indirect and direct comparison between array and standard Hall bar.** *****a****, Precision measurements of a 100 Ω resistance standard using both a standard Hall bar (blue) and two subarrays (Array1 red and Array2*

green). The 100 $\Omega$ is biased with 3 mA and the Hall bar and subarray receive 23 $\mu$A and 2.75 mA respectively. The top graph shows the different CCC readings, with one standard deviation error bars. The bottom graph shows the corresponding Allan deviations, with estimated relative error. The mean relative deviations are $\Delta_{HB\text{-}100}$ = (-4.4521 ± 0.0019) $\mu\Omega/\Omega$, $\Delta_{Array1\text{-}100}$ = (-4.4519 ± 0.0003) $\mu\Omega/\Omega$ and $\Delta_{Array2\text{-}100}$ = (-4.4522 ± 0.0005) $\mu\Omega/\Omega$, with standard error taken from Allan at $10^4$ s. **b**, Example direct comparison between Hall bar and one subarray, with standard error limited by Allan to ~ 0.2 n$\Omega/\Omega$. **c**, Mean relative deviation for direct comparison between Hall bar and array, calculated from precision measurements like in **b**. The purple data represent positive field direction, while green represent negative field direction. The error bars represent one standard error, taken from Allan deviation at $10^4$ s. The measurements were taken at three different field strengths (4.5, 5.0 and 5.5 T), but the data has been offset in the x-axis for clarity. **d**, Histogram of the data which produced the means in **c**. Each count represents one 20 minutes long CCC-reading. The distribution is normal and centered around 0.17 n$\Omega/\Omega$.

Finally, we explored the performance limits of the arrays in terms of bias current. Precision measurements (Fig. 3c) show that at least 5 mA is possible for sub-n$\Omega/\Omega$ precision, and deviations around 1 n$\Omega/\Omega$ are possible at currents up to 10 mA and 5 T (Fig. 4a and Supplementary 4). The quantization was tested by performing precision measurements at different fields (Fig. 4b, c). The apparent magnetic field dependence indicates that $I_C$ is at its limit for epigraphene (imperfect quantization), NbN contacts (resistive state), or a combination of both, since $I_C$ can improve at lower fields for either[4,26]. The deviation at 8.5 mA is within 1 n$\Omega/\Omega$ at lower magnetic fields < 5 T, which is acceptable for most practical metrological applications[8,24], including the Kibble balance[17]. Note that the fabrication techniques employed herein allow for further performance improvements. The observed $I_C$ is still far from any fundamental material limit and is simply restricted by the current device design. Since the NbN-leads can easily be made much larger (e.g. thicker film), what ultimately limits the QHA $I_C$ is the single graphene Hall bar $I_C$. By tuning the carrier density to a higher value[28], an array with $I_C$ > 10mA and good quantization should be achievable at 2 K and 5 T[8], and $I_C$ can be even higher under other operating conditions[3].

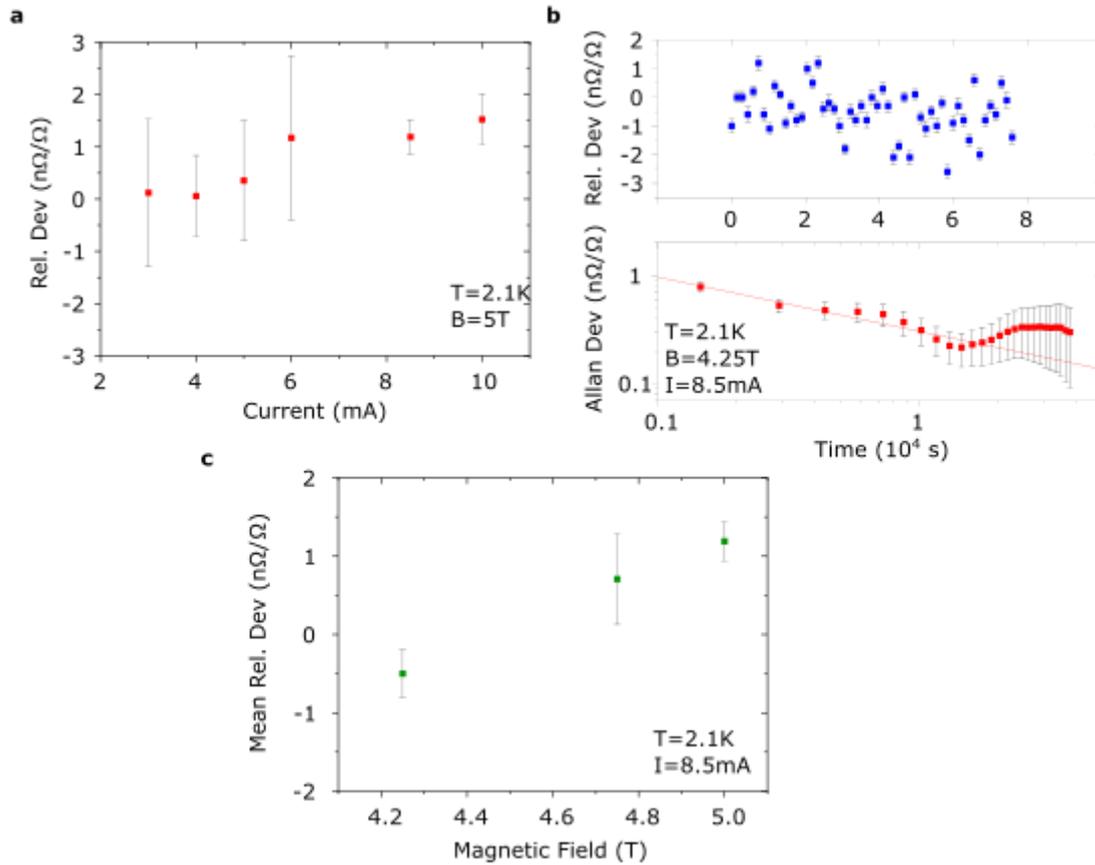

*Figure 4.* **High bias current measurements on arrays. a,** *CCC-measurements of a direct comparison between subarrays shows no significant deviation until 8.5 mA. The data consists of the mean of 5-10 CCC-readings (20 min long) so there is no Allan deviation analysis, and the error bars are therefore one standard deviation.* **b,** *Longer precision measurements with Allan deviation. The top graph shows the relative deviation, where each point is a 20 min long CCC-reading, with error bars of one standard deviation. The bottom graph shows the corresponding Allan deviation. The standard error is limited to 0.25 nΩ/Ω.* **c,** *Mean relative deviation calculated from measurements like those in* **b**. *The error bars represent one standard error, taken from Allan deviation at $10^4$ s. There appears to be a significant deviation at 5 T, which disappears into the uncertainty (k=2) at lower fields.*

In summary, we have demonstrated that a record-size 236-Hall bar graphene QHA is quantized with an unprecedented precision of 0.2 nΩ/Ω, opening the door for their application in metrology. The highest precision quantization remained up to at least 5 mA bias, with potential for operation at 8.5 mA and beyond. The proposed method of direct comparison of subarrays, coupled with reliable fabrication methods, paves the way for robust and flexible QHA designs with varied resistances which can be utilized in the new SI to decrease resistance calibration uncertainties, and help realize the electronic kilogram and quantum ampere. QHAs allow for the QHE to be more intimately involved in the improvement of the realization of several key units, and we believe that these arrays mark a new chapter for graphene in metrology and the SI.

## Acknowledgements

This work was jointly supported by VINNOVA (Ref. 2020-04311), the Swedish Foundation for Strategic Research (SSF) (Nos. GMT14-0077 and RMA15-0024), 2D TECH VINNOVA competence Center (Ref. 2019-00068), and Chalmers Excellence Initiative Nano. This work was performed in part at Myfab Chalmers.

## Author Contributions

K.C, T.B, H.H. and G.E. planned the experiments, H.H., N. S., S. L. and S. K. developed the fabrication processes, H.H. fabricated the graphene devices, H.H., K.C., T.B, and G.E. performed the electrical measurements and analyzed the data. H.H. and S.L. lead the writing process. All authors contributed to the writing process and reviewed the manuscript.

## Additional Information

Supplementary and Methods are available in the online version of the paper. Reprints and permissions information is available online at www.nature.com/reprints. Correspondence and requests for materials should be addressed to H.H.

## Competing Financial Interests

The authors declare no competing financial interests.

## Methods

The authors declare that all mentions of named products and companies are purely for reference and should not be taken as endorsements.

**Graphene growth**

Epigraphene chips (7x7 mm$^2$) were purchased from Graphensic AB. They were grown using thermal decomposition of silicon carbide[31] and had a monolayer coverage over 95%.

**Fabrication Methods**

For simplicity, the two subarrays were designed to consist only of parallel connections of individual Hall bars, with a single series connection between them. The number of parallel devices is 118 for each subarray, and this unusual resistance value of $h/236e^2$ was chosen because its ratio to 100 Ω is very close to 70/64, which is compatible with the winding ratios in the CCC[29]. The individual array Hall bar elements have a straight-forward minimalistic two-probe connection scheme in order to improve packing density, minimize complexity and increase device yield. Each hall bar element is contacted using a split contact, with six 15 μm wide leads spaced of 22 μm apart.

The devices were fabricated using standard electron beam lithography. Due to the nature of the chemical doping, only poly(methyl methacrylate) (PMMA) resists are suitable to contact the surface of graphene. The first lithography step was to make the NbN-contacts. A special three layer resist structure (to be published, see also PhD Thesis, He, H. *Molecular Doping of Epitaxial Graphene - For Device Applications* (2020)) was used, with PMMA directly on graphene (150 nm thick), followed by a copolymer poly(methylmethacrylate-co-methacrylic acid) (3000 nm) and finally with AR-P 6200 (200 nm) on top. The exposure dose was tuned in such a way that the NbN film can be properly anchored to SiC, while still being in direct electrical contact to graphene. After exposure and development, graphene underwent short reactive ion etching (RIE) with oxygen plasma (~ 30s) to expose some of the SiC underneath. Then 120 nm of NbN was sputtered in a magnetron system. The sample was then immediately transferred to an electron beam evaporator to deposit a 20 nm protective layer of Pt to prevent oxidation. For the second lithography step, a single layer PMMA (150 nm) was used as a mask to define the Hall structures using a longer RIE etching in oxygen plasma (~ 1 min).

After lithography, the sample was doped using chemical doping with F4TCNQ molecules[28]. This ensures a stable, homogenous, and controllable doping over the whole chip. We aimed to achieve a carrier density on the order of $10^{11}$ cm$^{-2}$ which is suitable for quantum Hall measurements around 2K and 5 T [4].

**Measurement Setup**

The devices were enclosed inside a dry TeslatronPT cryostat system, with a 12 T superconducting magnet and a base temperature of 1.5 K. The wiring consists of insulated copper leads with measured leakage resistance > 25 TΩ. The influence of this leakage on a resistor of 100 Ω (subarrays are ~ 109 Ω) is entirely negligible, but for a normal quantum Hall resistance standard it can lead to an error in the comparison measurements on the order of 0.1 nΩ/Ω or more. However, this small deviation is usually within the noise level of the CCC.

All measurements were conducted at a temperature around 2 K, which was measured using a Cernox thermometer mounted next to the chip carrier. For the precise CCC-measurements liquid helium was condensed inside the sample chamber and the sample was submerged in helium at 2.1 K, near the superfluid transition for the optimal temperature stability and maximal heat dissipation[32].

For regular measurements such as initial characterization, the samples were biased using a source (Keithley 6430A) and measured using a nanovoltmeter (Keithley 2182A). The measurement cables were twisted pairs copper leads with no significant additional shielding or filtering, and the noise level was limited to ~ 100 nV.

The precision measurements were performed inside a CCC-system from Oxford Instruments. It can very accurately compare two resistances by measuring their current ratio. For the comparison between Array1 and Array2 the winding ratios were set to Q = 64/64, for subarray versus 100 Ω standard Q = 70/64, Hall bar versus 100 Ω standard Q = 4130/32, and Hall bar versus 118x subarray Q = 3776/32. These ratios also determine the current ratio. The 100 Ω standard was always biased with 3 mA (limit due to heating), which automatically sets the current for the comparison QHE resistor (subarray or Hall bar) according to the resistance ratio. For Hall bar and array measurements, various current levels where used to check the critical current.

**Data analysis**

The CCC-system provides a measure of the resistance ratio Q = $R_B/R_A$ between two resistors A and B. This is then expressed as the relative deviation of test resistor B from its nominal value as referred to reference resistor A. For instance, the relative deviation $\Delta_{A-B}$ = (Q*$R_A$ -$R_{B,nominal}$)/ $R_{B,nominal}$. The value of $R_A$ is the reference value, and is usually chosen to be a fixed quantized resistance value. $R_{B,nominal}$ is the nominal value of resistor B, and $\Delta_{A-B}$ describes how much the measured value deviates from the nominal value.

All standard deviations and errors in this paper are stated as the error with unity coverage factor (k = 1), unless otherwise specified.

Where applicable, the mean relative deviations are presented as the weighted means of CCC-readings using variance weights taken from each reading[33]. The weighted mean of *n* samples of CCC-reading points $x_i$ with individual standard deviation $\sigma_i$ is:

$$\bar{x} = \frac{\sum_{i=1}^{n} x_i \sigma_i^{-2}}{\sum_{i=1}^{n} \sigma_i^{-2}}$$

With the standard error (variance) of the weighted mean:

$$\hat{\sigma}_{\bar{x}}^2 = \frac{1}{\sum_{i=1}^n \sigma_i^{-2}} \frac{1}{(n-1)} \sum_{i=1}^n \frac{(x_i - \bar{x})^2}{\sigma_i^2}$$

Note that the standard error is sometimes used instead of standard deviation as the weight when calculating the mean of several means. Unless specified otherwise, the standard error is usually directly taken from Allan deviation analysis (in the region of white noise) instead of using the equation above. In fact, the $n^{-1/2}$ scaling is motivated only when white noise dominates.

The Allan deviation reported in this paper is the overlapping Allan deviation. For CCC-reading data $x_i$, in total $N$ samples, taken with $\tau_0$ time difference, and $n$ readings in a bin, the Allan variance at time $n\tau_0$ is calculated as:

$$\sigma_A^2(n\tau_0, N) = \frac{1}{2n^2\tau_0^2(N-2n)} \sum_{i=0}^{N-2n-1} (x_{i+2n} - 2x_{i+n} + x_i)^2$$

The time difference $\tau_0$ is calculated as the average time difference between subsequent measurements. Each reading typically takes 20 min.

The relative error in for each point in the Allan deviation is estimated to be proportional to the inverse of the bin size $n$[34]:

$$err_{A\%} = \frac{1}{\sqrt{2(\frac{N}{n} - 1)}}$$

# Supplementary Information for "Exceptionally accurate large graphene quantum Hall arrays for the new SI"

Hans He[1], Karin Cedergren[1], Naveen Shetty[2], Samuel Lara-Avila[2,3], Sergey Kubatkin[2], Tobias Bergsten[1] and Gunnar Eklund[1].

1: RISE Research Institutes of Sweden, Box 857, S-50115 Borås, Sweden

2: Department of Microtechnology and Nanoscience, Chalmers University of Technology, 412 96 Gothenburg, Sweden

3: National Physical Laboratory, Hampton Road, Teddington TW11 0LW, United Kingdom

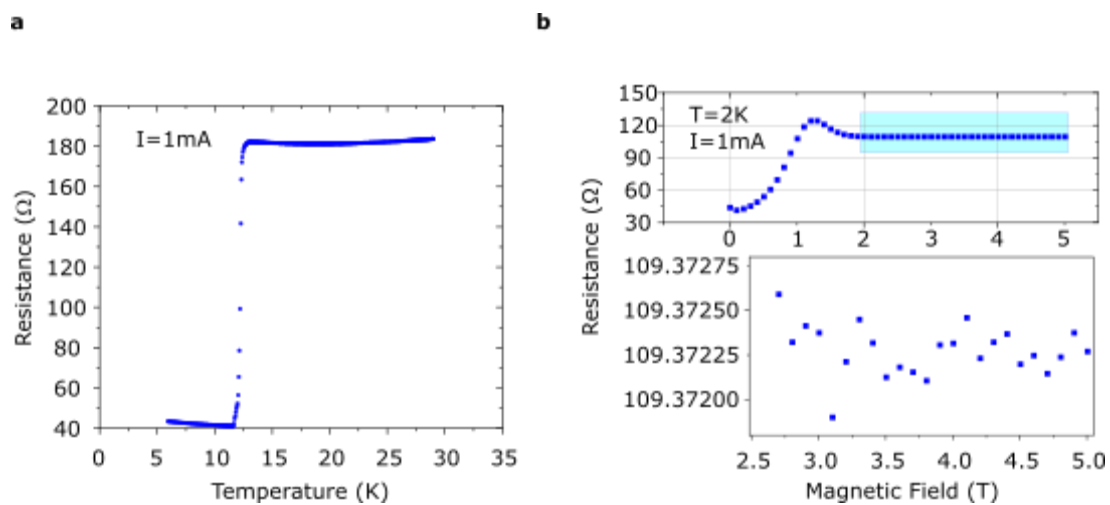

*Figure S1. **Subarray characterization**. **a**, shows the superconducting transition of the NbN-contacts (critical temperature $T_C$ = 12 K) measured for one subarray in configuration. The resistance increase after the superconducting phase transition is due to quantum effects in epigraphene[35]. **b**, shows the magnetotransport characterization of the same subarray, which appears fully quantized after 3 T. This demonstrates that the Hall bar and array have comparable carrier density and mobility. The offset of 4 mΩ (4 µV) from the quantized resistance $h/236e^2$ is due to voltmeter error.*

Note that due to its geometry, it is not possible to determine the carrier density or mobility of the subarray via regular Hall measurements. However, since its transition field into QHE it comparable to the Hall bar, both above 3 T, their electronic properties must also similar. This is to be expected since the molecular doping method produces homogenous doping, with carrier density differences within $10^{10}$ cm$^{-2}$ at 2 K.

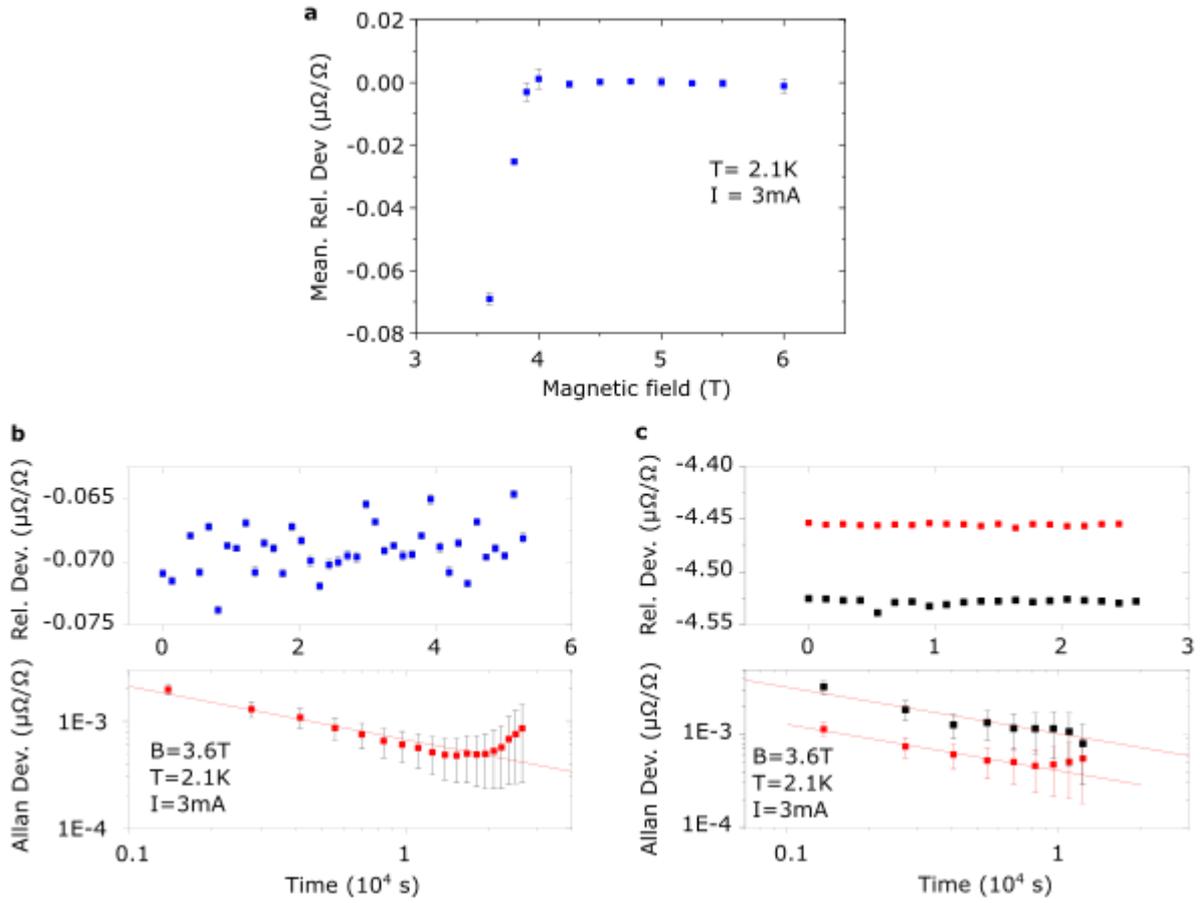

*Figure S2.* **CCC comparison measurements between array and Hall bar under non-quantizing conditions. a**, Mean relative deviation versus magnetic field for direct comparison of subarrays Array1 vs Array2. The data represent the average of a smaller collection of CCC-readings (20 min each) and the error bars represent one standard deviation (no Allan deviation). Proper quantization is clearly lost below 4 T and there is a significant deviation in resistance. **b**, Top graph shows CCC-readings for Array1 vs Array2 in non-quantizing state at 3.6 T field. The error bars represent one standard deviation. The bottom graph shows the corresponding Allan deviation. **c**, Top graph shows CCC-readings Array1 (black) and Array2 (red) vs 100 Ω standard. The bottom graph shows the corresponding Allan deviations.

Fig. S2a shows CCC-measurements taken at different magnetic fields, and they reveal that the quantization is lost below 4 T. By intentionally measuring the array in non-quantizing state a significant deviation can be produced, and this can be used as an additional test of the comparison measurements.

Fig. S2b shows the direct comparison of Array1 and Array2 at 3.6 T. In this non-quantized state, the subarrays differ significantly, and has relative deviation of $\Delta_{Arra1-Array2}$ = (-0.0690 ± 0.0006) µΩ/Ω. The error denotes the standard error of the mean, extracted from the corresponding Allan deviation.

Fig. S2c shows a similar comparison of each subarray to a 100 Ω standard. Array1 versus 100 Ω has a relative devation of $\Delta_{Arra1-100\Omega}$ = (-4.5280 ± 0.0011) µΩ/Ω. The error denotes the standard error of the mean, extracted from the corresponding Allan deviation. Because this subarray is not quantized, the measured value of the 100 Ω standard differs from its nominal value, which should correspond to a relative deviation of around -4.452 µΩ/Ω as in Fig. 2a in the main text. On the other hand, Array2 versus 100 Ω has a relative deviation of $\Delta_{Arra2-100\Omega}$ = (-4.4550 ± 0.0004) µΩ/Ω. The error denotes the standard error of the mean, extracted from the corresponding Allan deviation. This is much closer to the measurement in quantizing conditions. This means that Array1 loses its quantization before Array2, and the indirect comparison has the deviation $\Delta_{Arra1\text{-}Array2,\ indir}$ = (-0.0730 ± 0.0013) µΩ/Ω, which agrees well with the direct comparison $\Delta_{Arra1\text{-}Array2}$ = (-0.0690 +/- 0.0006) µΩ/Ω. The reason that Array1 and Array2 have different quantizing fields is likely due to slight difference in carrier density and mobility. The molecular doping method, while generally homogenous, can yield a finite doping difference on the order of $10^{10}$ cm$^{-2}$ [28]. Beside the difference in quantizing field, once can also reasonably expect a difference in critical current at a given field between the two subarrays.

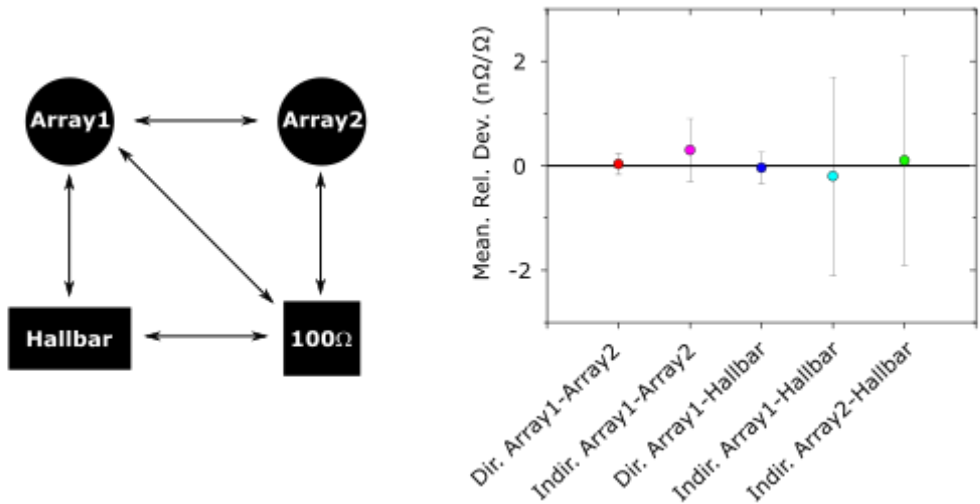

*Figure S3. **Summary of comparison measurements.*** The arrows in the left diagram depict direct comparisons between different resistance standards. Both direct and indirect comparisons for the subarrays show no significant deviation. The error bars are one standard error, limited by Allan deviation measurements.

In the main test we have demonstrated agreement between different combinations of direct and indirect comparisons between a quantized Hall bar and the subarrays. These measurements are summarized in Fig. S2, and the weighted mean of all such comparisons is $\Delta_{Total}$ = 0.03 +/- 0.04 nΩ/Ω, which is zero within the uncertainty. The consistency of the comparison measurements can be checked by looking at the three closed comparison loops. Inside each loop, the relative deviations should sum to zero. For instance, $\Delta_{Array1\text{-}Array2}$ + $\Delta_{Array2\text{-}100}$ + $\Delta_{100\text{-}Array1}$ = $\Delta_{Array1\text{-}Array2}$ + $\Delta_{Array2\text{-}100}$ - $\Delta_{Array2\text{-}100}$ = 0.033 +/- 0.62 nΩ/Ω, which is zero within the expanded measurement uncertainty. The other two loops are dominated by the uncertainty of the measurement $\Delta_{HB\text{-}100}$ and are zero well within an expanded uncertainty of 2 nΩ/Ω.

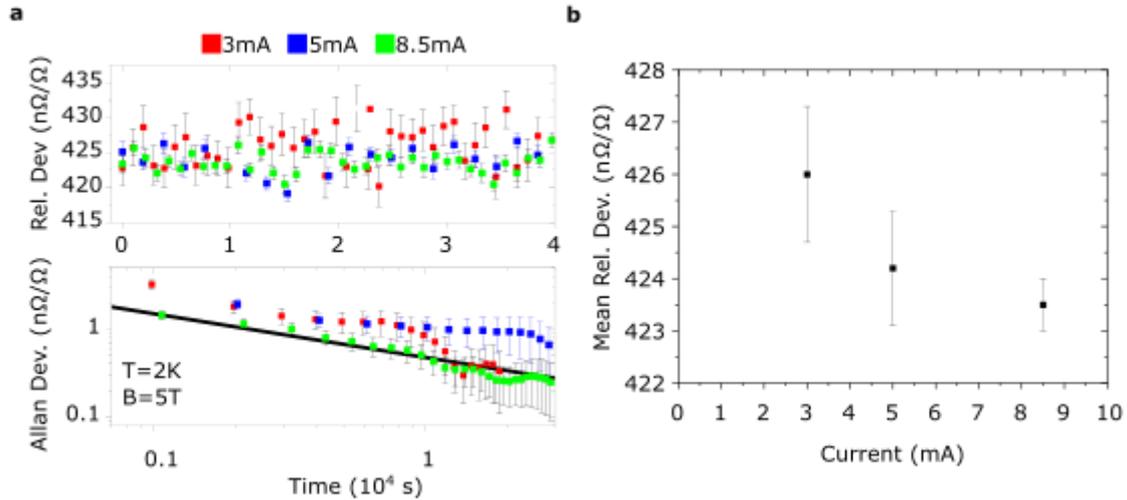

*Figure S4. **Comparison measurements between one subarray and a 12.9 kΩ standard. a,** Collection of CCC-measurements taken at different bias currents, with corresponding Allan deviations. The solid black line shows a $1/\tau^{1/2}$ trend for white noise. **b,** The weighted mean of the data in **a**, with error bars being one standard error of the mean taken directly from Allan deviations.*

In the main test we have shown that at 8.5 mA bias and 5 T there appears to be a slight deviation of around 1 nΩ/Ω between the two subarrays. We wish to demonstrate that this deviation is due to a small deviation from perfect quantization on the order of 1 nΩ/Ω in one of the subarrays, and that it is not the case that they have both deviated very far from quantized conditions in unison. To achieve this, we performed comparison measurements between one subarray and a standard resistor with a nominal value of 12.9 kΩ, kept in a temperature-controlled air bath. This standard is used instead of the 100 Ω standard described in the main text because it can withstand higher currents. CCC-measurements were performed at different bias currents and the data are summarized in Fig. S3(b). We see that there is no significant change in the relative deviation between bias currents, and it is all within the expanded measurement uncertainty of 3.4 nΩ/Ω (k = 2) for the deviation of 3 mA compared to 5 mA, and 2.4 nΩ/Ω (k = 2) for 5 mA compared to 8.5 mA, and 2.8 nΩ/Ω (k = 2) for 3 mA compared to 8.5 mA. While this measurement is much noisier than the direct subarray versus subarray measurements, it still shows that the quantization of the individual subarray changes at most a few parts-per-billion, limited by noise due to the external 12.9 kΩ standard. For comparison, the precision measurements for the direct comparison between subarray versus subarray, and subarray versus Hall bar, show that the quantization is 0.4 nΩ/Ω (k = 2) for currents 3 mA and 5 mA. This test supports the notion that a deviation on the order of 1 nΩ/Ω develops occurs in one of the two subarrays at high bias currents, similarly to how one subarray loses quantization at lower fields before the other. This could be attributed once again to slight differences in electrical properties such as carrier density.